\documentclass[aps,pre,reprint,superscriptaddress,showpacs,showkeys]{revtex4-1} 
\usepackage{amsfonts}
\usepackage{amssymb}
\usepackage{amsmath}
\usepackage{graphicx}
\usepackage{color}

\begin{document} 

\title{Statistical properties of Barkhausen noise in amorphous ferromagnetic films} 
\author{F.~Bohn} 
\email[Electronic address: ]{felipebohn@gmail.com}
\affiliation{Escola de Ci\^{e}ncias e Tecnologia, Universidade Federal do Rio Grande do Norte, 59078-970 Natal, RN, Brazil} 
\affiliation{Departamento de F\'{i}sica Te\'{o}rica e Experimental, Universidade Federal do Rio Grande do Norte, 59078-970 Natal, RN, Brazil} 
\author{M.~A.~Corr\^{e}a} 
\affiliation{Departamento de F\'{i}sica Te\'{o}rica e Experimental, Universidade Federal do Rio Grande do Norte, 59078-970 Natal, RN, Brazil} 
\author{M.~Carara} 
\affiliation{Departamento de F\'{i}sica, Universidade Federal de Santa Maria, 97105-900 Santa Maria, RS, Brazil}
\author{S.~Papanikolaou} 
\affiliation{Department of Mechanical Engineering and Materials Science and Department of Physics, Yale
University, 06520-8286 New Haven, Connecticut, USA} 
\author{G.~Durin} 
\affiliation{INRIM, Strada delle Cacce 91, 10135 Torino, Italy} 
\affiliation{ISI Foundation, Viale S.\ Severo 65, 10133 Torino, Italy} 
\author{R.~L.~Sommer} 
\affiliation{Centro Brasileiro de Pesquisas F\'{i}sicas, Rua Dr.\ Xavier Sigaud 150, Urca, 22290-180 Rio de Janeiro, RJ, Brazil} 

\date{\today} 

\begin{abstract} 
We investigate the statistical properties of the Barkhausen noise in amorphous ferromagnetic films with thicknesses in the range between $100$ and $1000$ nm. From Barkhausen noise time series measured with the traditional inductive technique, we perform a wide statistical analysis and establish the scaling exponents $\tau$, $\alpha$, $1/\sigma \nu z$, and $\vartheta$. We also focus on the average shape of the avalanches, which gives further indications on the domain wall dynamics. Based on experimental results, we group the amorphous films in a single universality class, characterized by scaling exponents $\tau\sim 1.27$, $\alpha \sim 1.5$, $1/\sigma \nu z \sim \vartheta \sim 1.77$, values similar to that obtained for several bulk amorphous magnetic materials. Besides, we verify that the avalanche shape depends on the universality class. By considering the theoretical models for the dynamics of a ferromagnetic domain wall driven by an external magnetic field through a disordered medium found in literature, we interpret the results and identify an experimental evidence that these amorphous films, within this thickness range, present a typical three-dimensional magnetic behavior with predominant short-range elastic interactions governing the domain wall dynamics. Moreover, we provide experimental support for the validity of a general scaling form for the average avalanche shape for non-mean-field systems.
\end{abstract} 



\maketitle 

\section{Introduction} 
\label{Introduction}

Barkhausen noise (BN) can be understood as a result of the complex microscopic magnetization process and irregular motion of domain walls (DWs) in ferromagnetic materials~\cite{Resumao_BN}. In recent years, it has attracted growing interest as one of the best examples of response of a dynamical disordered system exhibiting crackling noise, becoming an excellent candidate for investigating scaling phenomena~\cite{N410p242, NP1p13, NP3p518, Resumao_francesca}. From this new point of view, BN becomes an important playground for investigations since several systems in many situations, remarkably, present response signals, or time series, that share common characteristic features. This is the case of, besides BN in ferromagnetic materials~\cite{Resumao_francesca, NP7p316, PRE86p066117, PRE88p032811}, the seismic activity in earthquakes~\cite{RG34p433, PRL78p4885, PRE73p056104}, the dynamics of vortices in supercondutors~\cite{PRL73p1703, PRL74p1206, PRB53p3520}, the fluctuations in the stock market~\cite{PA246p430,C15p026104}, the acoustic emission in micro-fractures processes~\cite{N388p658, PRL89p185503}, the shear response of a granular media~\cite{PRL96p118002, NP7p554}, and propagation of fluids in porous media~\cite{Fractal_concepts}. The reason why the interest is revived in this classical and old effect is mainly motivated by a fundamental question whether there is any simple law governing the seemingly random avalanches events. 

Noise statistical analysis suggests that general systems with avalanche dynamics can be classified into universality classes characterized by the values of the scaling exponents~\cite{NC4p2927}. An universality class of the Barkhausen noise in a sample is commonly identified by measuring the distributions of Barkhausen avalanche sizes and durations, average avalanche size as a function of its duration, and power spectrum, which, typically, display scaling in a quite large range, with critical exponents $\tau$, $\alpha$, $1/\sigma \nu z$, and $\vartheta$, respectively~\cite{Resumao_BN}. In particular, the statistical properties seem to exhibit universality, i. e., they are independent on the microscopic details of the dynamics, being controlled only by general properties such as the system dimensionality and the range of the relevant interactions~\cite{Fractal_concepts}.

Experimentally, large efforts have been devoted to relate the scaling exponents to general properties of the avalanche dynamics. For bulk materials, such as ribbons and sheets, well-known by exhibiting three-dimensional magnetic behavior, there is an established and consistent interpretation of the BN statistical properties. Despite the large number of experimental works~\cite{PRL67p1334, PRE50p3446, JAP68p2908, PRL75p276, PRE54p2531, PRE59p3884, PRE65p046139, JMMM140p1835, PRL84p4705}, for a long time, the universality seemed difficult to be confirmed since the experimental exponents spread in a wide range of values and did not show a good agreement with theoretical results. However, nowadays, the results are understood in terms of the depinning transition of domain walls with two distinct universality classes for amorphous and polycrystalline materials, associated to distinct ranges of the interactions governing the DWs dynamics~\cite{PRL84p4705}.

It is noticeable that most of the studies reported so far are related to three-dimensional systems and bulk samples. For two-dimensional systems and samples with reduced dimensions, the BN statistical properties are less clear. On the theoretical side, models and simulations~\cite{PRL84p1316, JSMP08020, IEEETM46p228, PA390p4192, PRE64p066127, PRE69p026126, PRL50p1486, PRE53p414, PRB51p6296, JPCM14p2353} infer the existence of two distinct universality classes, according the range of interactions governing the DWs dynamics, as well as indicate that three and two-dimensional systems present distinct exponents. 

Experimentally, several interesting results have been obtained for different ferromagnetic films through both magneto-optical~\cite{PRL84p5415, IEEETM36p3090, JAP101p063903, PRL90p0872031, JMMM310p2599, JAP93p6563, NP3p547, JAP103p07D907, SSC150p1169, JAP109p07E101, PRB83p060410R} and inductive~\cite{PB384p144, NP7p316, PRE88p032811} techniques. With the first one, E.~Puppin {\it et al.} have reported the exponent $\tau \sim 1.1$ for Fe crystalline films, with thickness of $90$ nm~\cite{PRL84p5415, IEEETM36p3090}, and, recently, have estimated $\tau \sim 0.8$ - $1.2$ for amorphous Fe$_{73.5}$Cu$_1$Nb$_3$Si$_{22.5}$B$_4$ ferromagnetic films, with variable thickness between $25$ and $1000$ nm~\cite{JAP101p063903}, both results obtained through measurements performed using a magneto-optical elipsometer. D.-H.~Kim {\it et al.}~\cite{PRL90p0872031} have presented $\tau \sim 1.33$ for Co polycrystalline films, with thicknesses varying from $5$ to $50$ nm, and S.-C.~Shin {\it et al.}~\cite{JMMM310p2599} have found $\tau \sim 1.33$ for Co and MnAs films, with the same thicknesses, by measurements using a magneto-optical microscope magnetometer, capable of observing directly the motion of the DWs and the Barkhausen avalanches~\cite{JAP93p6563}. Following the same line, K.-S.~Ryu {\it et al.}~\cite{NP3p547, JAP103p07D907} have shown for a $50$ nm-thick MnAs film the crossover between two distinct universality classes, which is caused by the competition between long-range dipolar interaction and the short-range DW surface tension, with $\tau$ varying from $1.32$ to $1.04$, tuned by an increase of temperature from $20^\circ$C until $35^\circ$C. More recently, with similar experiments, S.~Atiq {\it et al.}~\cite{SSC150p1169} have obtained $\tau \sim 1.02$ for $\gamma$-Fe$_4$N films irrespective of the degree of texture of the film induced by annealing temperature, while H.-S.~Lee have found $\tau \sim 1.1$ for $50$ nm thick Ni$_x$Fe$_{1-x}$ films, with $x=0$, $0.3$, $0.4$ and $0.5$, independent of the film composition~\cite{JAP109p07E101}, and $\tau\sim 1.33$ for NiO($t_\textrm{NiO}$)/Fe($30$ nm) films with $t_\textrm{NiO} = 0$, $30$, $80$, $100$, and $150$ nm~\cite{PRB83p060410R}. Considering these reports found in literature based on magneto-optical measurements~\cite{PRL84p5415, IEEETM36p3090, JAP101p063903, PRL90p0872031, JMMM310p2599, JAP93p6563, NP3p547, JAP103p07D907, SSC150p1169, JAP109p07E101, PRB83p060410R}, although they restrict the analysis to distributions of jump sizes, when compared to theoretical predictions, most of them does confirm an essentially two-dimensional magnetic behavior for films, as expected due to reduced thickness of the studied samples. 

On the other side, via the traditional inductive technique~\cite{PB384p144, NP7p316, PRE88p032811}, our group has reported results suggesting that the two-dimensional magnetic behavior commonly verified for films cannot be generalized for all thickness ranges. In this sense, L.~Santi {\it et al.}~\cite{PB384p144} have presented exponents $\tau \sim 1.25$ and $\alpha \sim 1.6$ for amorphous Fe$_{73.5}$Cu$_1$Nb$_3$Si$_{22.5-x}$B$_x$, with $x = 4$ and $9$, ferromagnetic films in a wide range of thickness. S.~Papanikolaou {\it et al.}~\cite{NP7p316} have obtained a wide BN statistical analysis for a $1000$ nm-thick Permalloy polycrystalline ferromagnetic film and verified driving rate-dependent $\tau$ and $\alpha$, while $1/\sigma \nu z$ and $\vartheta$ constant critical exponents. Moreover, as a fundamental signature of the avalanches, the average temporal avalanche shape has been estimated and shown to be given by a symmetric inverted parabola, providing strong indications of the underlying physics, such as the system dimensionality and kind and range of interactions governing the avalanche dynamics. More recently, we have obtained the same wide statistical analysis for Permalloy polycrystalline ferromagnetic films with thicknesses between $100$ and $1000$ nm~\cite{PRE88p032811}. In that case, we grouped the films in a single universality class since all films irrespective on the thickness are characterized by the scaling exponents $\tau \sim 1.5$, $\alpha \sim 2.0$ and $1/\sigma\nu z \sim \vartheta \sim 2.0$ at the smallest magnetic field rate. Thus, by considering the two latter reports, we identify an universal three-dimensional magnetic behavior, with predominant strong long-range dipolar interactions governing the domain wall dynamics, revealed by the quantitative agreement between experimental results and the well-known predictions for bulk polycrystalline magnets~\cite{PRL79p4669, PRB58p6353, PRL84p4705, PRE69p026126, PRE64p066127}.

In this paper we report an experimental evidence for a three-dimensional magnetization dynamics, governed by short-range elastic interactions of the DWs, in amorphous ferromagnetic films having different thickness from $100$ to $1000$ nm. We investigate the statistical properties of Barkhausen noise time series measured with the traditional inductive method. By applying the traditional statistical treatment employed for bulk materials, we analyze the scaling behavior in the distribution of Barkhausen avalanche sizes, the distribution of avalanche durations, the average avalanche size as a function of its duration, and the power spectrum and establish the exponents $\tau$, $\alpha$, $1/\sigma \nu z$, and $\vartheta$. Here, we go beyond power-laws and also focus on the average shape of the avalanches due to the irregular and irreversible motion of the domain walls in a disordered ferromagnet, verifying that avalanche shape depends on the universality class. Through this wide statistical analysis and the comparison to theoretical predictions, we group the amorphous films with distinct thicknesses in a single universality class, providing further information on the DWs dynamics in systems with reduced dimensions and the role of structural character and film thickness on the scaling behavior in the BN statistical properties in ferromagnetic films. Moreover, we provide experimental support for the validity of a general scaling form for the average avalanche shape for non-mean-field systems.

\section{Experiment} 
\label{Experiment}

For the study, we analyse experimental Barkhausen noise time series measured in a set of FeSiB amorphous ferromagnetic films with nominal composition Fe$_{75}$Si$_{15}$B$_{10}$ and thicknesses of $100$, $150$, $200$, $500$, and $1000$ nm. The films are deposited by magnetron sputtering onto glass substrates, covered with a $2$ nm thick Ta buffer layer. The deposition process is performed with the following parameters: base vacuum of $1.5 \cdot 10^{-7}$ Torr, deposition pressure of $5.2 \cdot 10^{-3}$ Torr with a $99.99$\% pure Ar at $20$ sccm constant flow, and DC source with current of $50$ mA and $65$ W set in the RF power supply for the deposition of the Ta and FeSiB layers, respectively. During the deposition, the substrate with dimensions of $10$ mm $\times$ $4$ mm moves at constant speed through the plasma to improve the film uniformity, and a constant magnetic field of $1$ kOe is applied along the main axis of the substrate during the film deposition in order to induce a magnetic anisotropy and define an easy magnetization axis. X-ray diffraction results calibrate the sample thicknesses and verify the amorphous structural character of all films. Quasi-static magnetization curves are obtained with a vibrating sample magnetometer, measured along and perpendicular to the main axis of the films, in order to verify the magnetic behavior.

The Barkhausen noise in ferromagnetic materials corresponds to the time series of voltage pulses detected by a sensing coil wound around a ferromagnetic material submitted to a slow-varying magnetic field~\cite{Resumao_BN, Resumao_francesca, PZ20p401, Bertotti}. The noise is produced by sudden and irreversible changes of magnetization, mainly due to the irregular motion of the domain walls in a disordered magnetic material, a result of the interactions between the DWs and pinning centers, such as defects, impurities, dislocations, and grain boundaries~\cite{Bertotti, Resumao_BN, Cullity, JMMM23p136, JMMM317p20}. 

We record Barkhausen noise time series using the traditional inductive technique in an open magnetic circuit. Sample and pickup coils are inserted in a long solenoid with compensation for border effects, to ensure an homogeneous applied magnetic field on the sample. The sample is driven by a $50$ mHz triangular magnetic field, applied along the main axis, with an amplitude high enough to saturate it magnetically. BN is detected by a sensing coil ($400$ turns, $3.5$ mm long, $4.5$ mm wide, and $1.25$ MHz resonance frequency) wound around the central part of the sample. A second pickup coil, with the same cross section and number of turns, is adapted in order to compensate the signal induced by the magnetizing field. The Barkhausen signal is then amplified, filtered, and, finally, digitalized. All BN measurements are performed under similar experimental conditions: $100$ kHz low-pass filter set in the preamplifier and signal acquisition with sampling rate of $4$ million samples per second. The time series are acquired just around the central part of the hysteresis loop, near the coercive field, where the domain wall motion is the main magnetization mechanism~\cite{Bertotti, PRB58p6353, JMMM317p20} and the noise achieves the condition of stationarity~\cite{JSMp01002}. 

The Barkhausen noise statistical properties are measured following the procedure discussed in detail in Refs.~\cite{JMMM140p1835, F3p351, PRL84p4705, NP7p316, PRE88p032811}. For each experimental run, the statistical properties are obtained from $150$ measured Barkhausen noise time series, by averaging the distributions over a $10^5 - 10^6$ avalanches. First of all, due to the reduced intensity of the signal, we employ a Wiener deconvolution~\cite{NP7p316}, which optimally filters the background noise, removes distortions introduced by the response functions of the measurement apparatus in the original voltage pulses, and provides us reliable statistics. Having established a sophisticated method of extraction of the BN avalanches and by considering a threshold value $v_r$ to properly define the beginning and end of each Barkhausen avalanche, the universality class of the Barkhausen noise is characterized primarily by measuring the distributions of Barkhausen avalanche sizes $(P(s))$ and durations $(P(T))$, the average avalanche size as a function of its duration $(\langle s(T) \rangle\, vs.\, T)$, and the power spectrum $(S(f))$. 

We observe that the measured $P(s)$, $P(T)$, and $\langle s(T) \rangle\, vs.\, T$ avalanche distributions follow a cuttoff-limited power-law behavior and they can be, respectively, fitted as~\cite{KDahmen_unpublished}
\begin{equation}
P(s) \sim s^{-\tau}e^{-(s/s_0)^{n_s}},
\label{eq_01}
\end{equation}
\begin{equation}
P(T) \sim T^{-\alpha}e^{-(T/T_0)^{n_T}},
\label{eq_02}
\end{equation}
\begin{equation}
\langle s(T) \rangle \sim T^{1/\sigma \nu z}\left[\frac{1}{1+(T/T_0)^{n_{ave}(1/\sigma \nu z-1)}}\right]^{1/n_{ave}},
\label{eq_03}
\end{equation}
where $s_0$ and $T_0$ indicate the position where the function deviates from the power-law behavior, and $n_s$, $n_T$, and $n_{ave}$ are fitting parameters related to the shape of the cutoff function. Here, the analysis of the statistical properties is done with the software BestFit~\cite{bestFit}, which allows us to fit them at the same time, respecting a well-known scaling relation between the exponents~\cite{PRB58p6353, PRB52p12651},
\begin{equation}
\alpha = (\tau-1)/\sigma \nu z + 1.
\label{eq_04}
\end{equation}
Although the power spectrum has not been considered for the fitting procedure, we observe that the measured $S(f)$ also follows a power-law behavior at the high frequency part of the spectrum and it can be described by~\cite{JSMp01002}
\begin{equation}
S(f) \sim f^{-\vartheta}.
\label{eq_05}
\end{equation}

Moreover, we also focus the analysis on the measurement of the average avalanche shape, a sharper tool for comparison between theory and experiments, going far beyond power laws, and being more informative than slopes~\cite{N410p242, Resumao_BN, PRE65p046139, PRB62p11699, Resumao_francesca, JMMM242p1085, NP1p46, NC4p2927}. The average avalanche shape has been estimated for a wide variety of systems, as planar crack front propagation experiments~\cite{PRL96p045501, PRE83p046108}, plastically deforming crystals~\cite{PRE74p066106}, earthquakes~\cite{PRE73p056104} and Barkhausen noise~\cite{PRE65p046139, NP1p46, NP7p316}. Here, we consider the average temporal avalanche shape, considering all the avalanches of a given duration $T$ and averaging the voltage signal at each time step $t$, as well as the average avalanche shape for a given size $s$ or magnetization, when considering all the avalanches of a size as and averaging the BN signal at each size step $S$. 

Recently, L.\ Laurson {\it et al.}~\cite{NC4p2927} have suggested that the general scaling form for the average temporal avalanche shapes for non-mean-field systems can be described by
\begin{equation}
\langle V(t|T)\rangle \propto T^{1/\sigma \nu z -1}\left [ \frac{t}{T} \left (1- \frac{t}{T} \right ) \right ]^{1/\sigma \nu z -1}\, ,
\label{eq_06}
\end{equation}
\noindent in which the exponent $1/\sigma \nu z$ is considered, resulting in a consequent evolution of the average avalanche shape with the universality class. Similarly, the general scaling form to avalanches of a given size can be written as
\begin{equation}
\langle V(S|s)\rangle \propto s^{1- \sigma \nu z}\left [ \frac{S}{s} \left (1- \frac{S}{s} \right ) \right ]^{1-\sigma \nu z }\, ,
\label{eq_07}
\end{equation}
\noindent suggesting similar dependance.

\section{Results and discussion}
\label{Results_and_discussion}

Here we show why the studied amorphous films present a typical three-dimensional magnetic behavior, with predominant short-range elastic interactions governing the domain wall dynamics.

\subsection{Structural and quasi-static magnetic characterization of the films} 

First of all, we characterize the films from the structural and quasi-static magnetic point of view. 

\begin{figure}[!ht] 
\includegraphics[width=8.5cm]{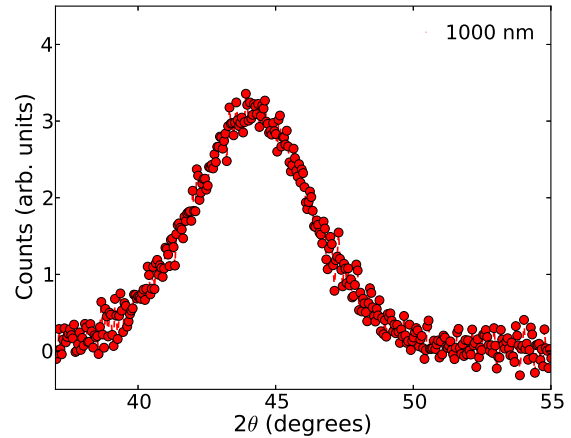}
\vspace{-.3cm}\caption{(Color online) High angle x-ray diffraction pattern for the FeSiB film with thickness of $1000$ nm. The diffraction pattern confirms the amorphous character of the film. The films with distinct thicknesses present similar behavior.} 
    \label{Fig_01} 
\end{figure}

Figure~\ref{Fig_01} shows an high angle x-ray diffraction pattern for the FeSiB film with thickness of $1000$ nm. For the films with distinct thicknesses, similar behavior is obtained. In particular, the pattern clearly indicates the amorphous state of the film, as depicted from the broad peak with low intensity, around $2\theta \sim 44^\circ$, and the absence of thin peaks with high intensity.

\begin{figure}[!ht] 
\hspace{-.15cm}\includegraphics[width=4.45cm]{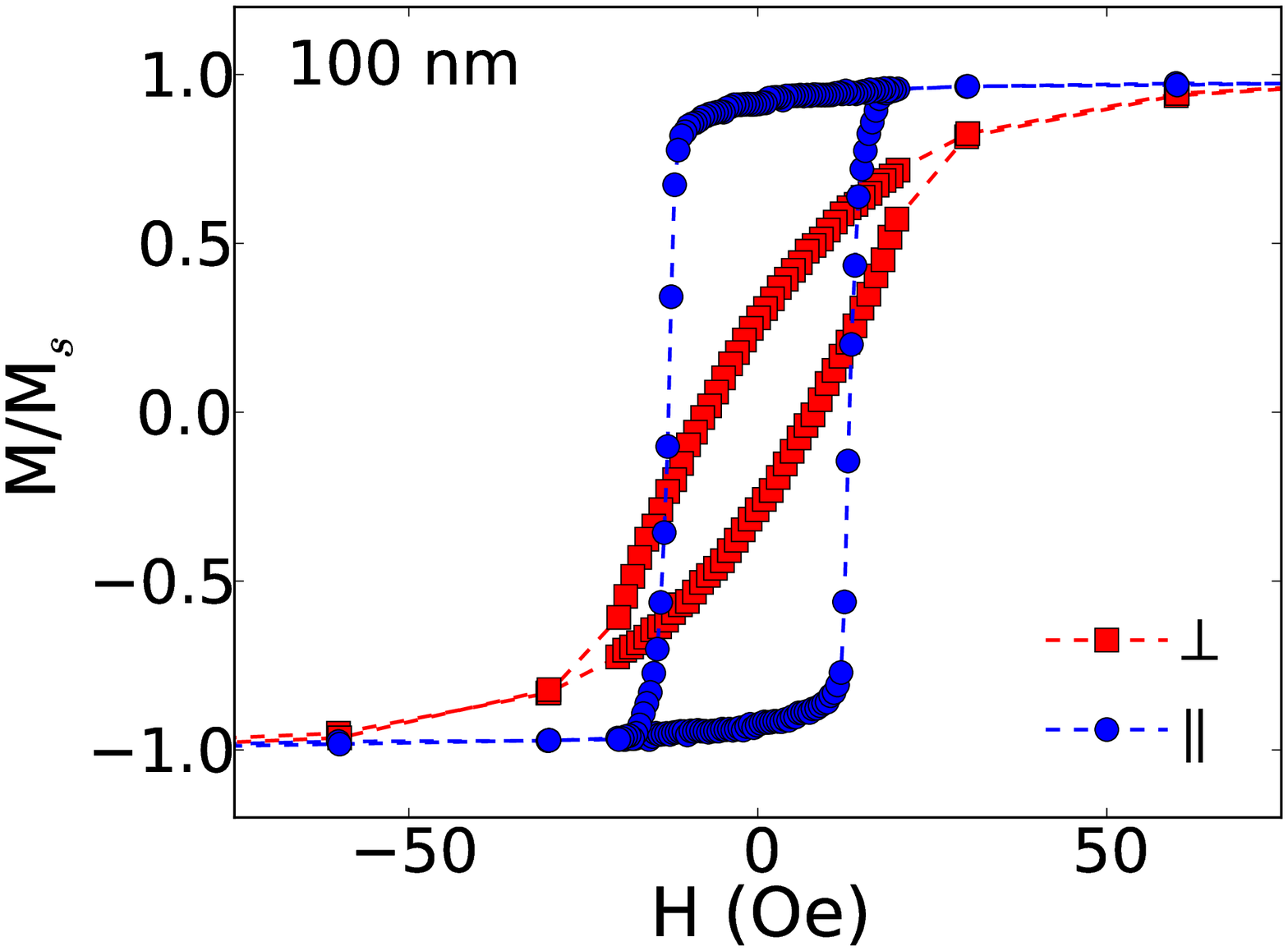} 
\hspace{-.25cm}\includegraphics[width=4.45cm]{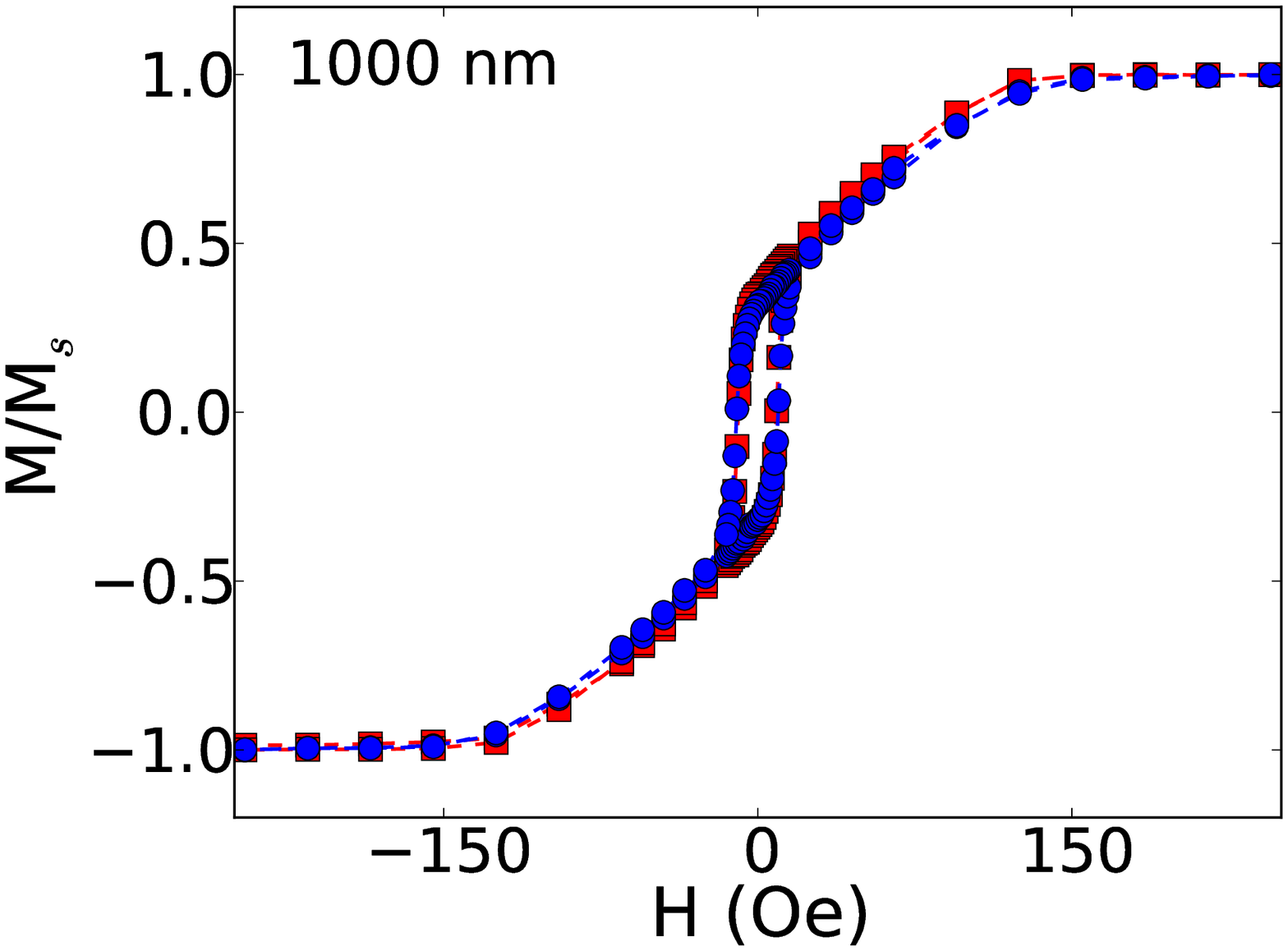}
\vspace{-.7cm}\caption{(Color online) Normalized quasi-static magnetization curves for the FeSiB films with the thicknesses of $100$ and $1000$ nm, obtained with the in-plane magnetic field applied along ($\parallel$) and perpendicular ($\perp$) to the main axis of the films. The change of magnetic behavior is observed in the thickness range between $200$ and $500$ nm.} 
    \label{Fig_02} 
\end{figure}

Figure~\ref{Fig_02} shows the quasi-static magnetization curves, measured with the in-plane magnetic field applied both along and perpendicular to the main axis, obtained for the FeSiB films with selected thicknesses. When analyzed as a function of the thickness, the magnetization curves indicate the existence of a thickness range, between $200$ and $500$ nm, which splits the films in two groups according the magnetic behavior, similarly to the one observed and discussed in details in Refs.~\cite{Dominios_magneticos, PB384p144, JAP101p033908, JAP103p07E732, JAP104p033902, JPDAP43p295004, PRE88p032811}. In this case, for films with thicknesses below $200$ nm, the angular dependence of the  magnetization curves indicates an uniaxial in-plane magnetic anisotropy, induced by the magnetic field applied during the deposition process. However, for films above $500$ nm, the curves exhibit isotropic in-plane magnetic properties, with an out-of-plane anisotropy contribution, a behavior related to the stress stored in the film and/or to a columnar microstructure as the thickness is increased.

\subsection{Barkhausen noise and statistical properties}

Here, as a fingerprint of the crackling noise in magnetic systems, the response of a ferromagnetic system to a smooth and slow external magnetic field is the Barkhausen noise, characterized by a series of discrete and irregular avalanches, due to sudden and irreversible jumps of the magnetization, with a broad range of sizes and duration times, separated by quiescent time intervals. 

\begin{figure*}[!] 
\centering
\includegraphics[width=8.5cm]{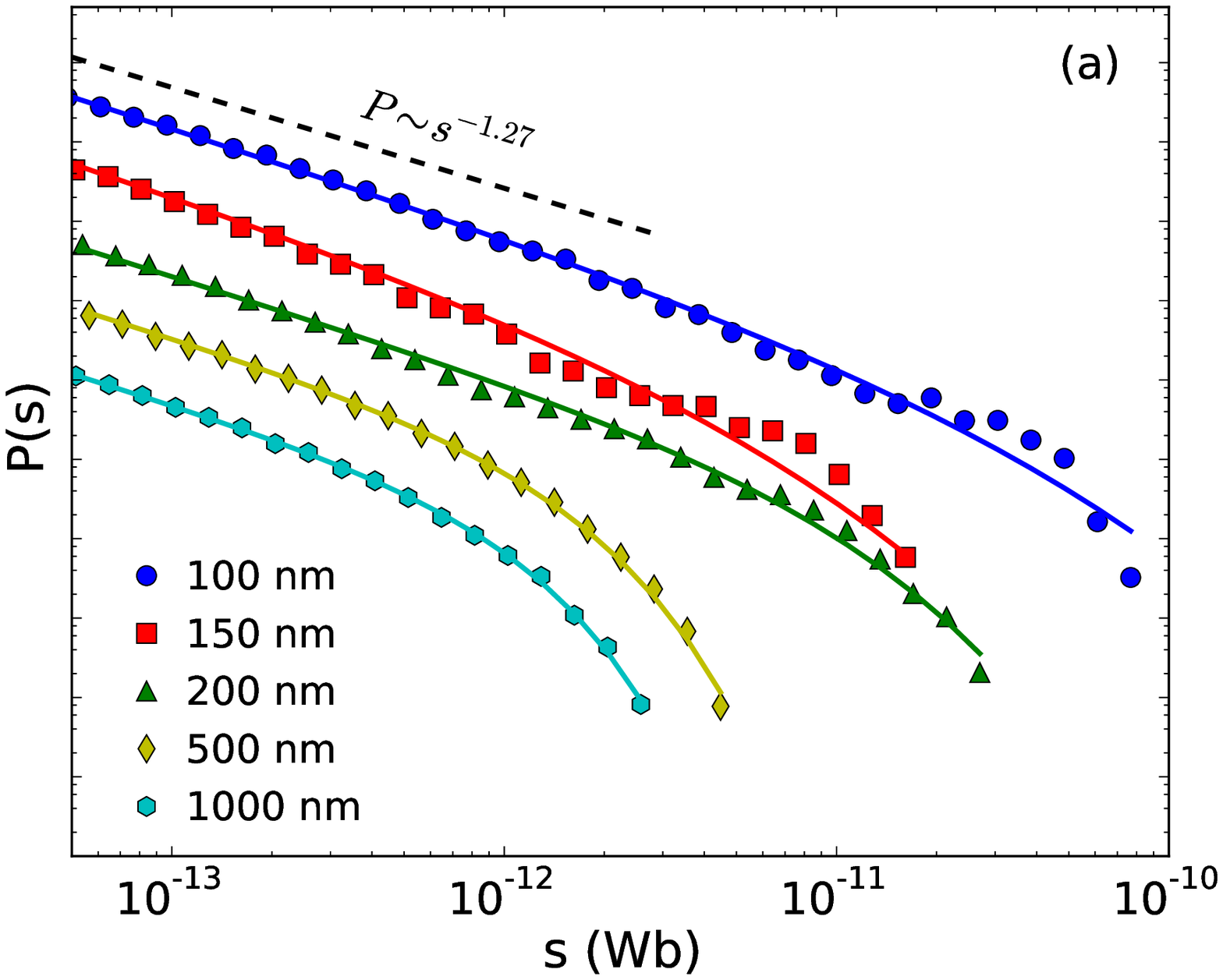}
\includegraphics[width=8.5cm]{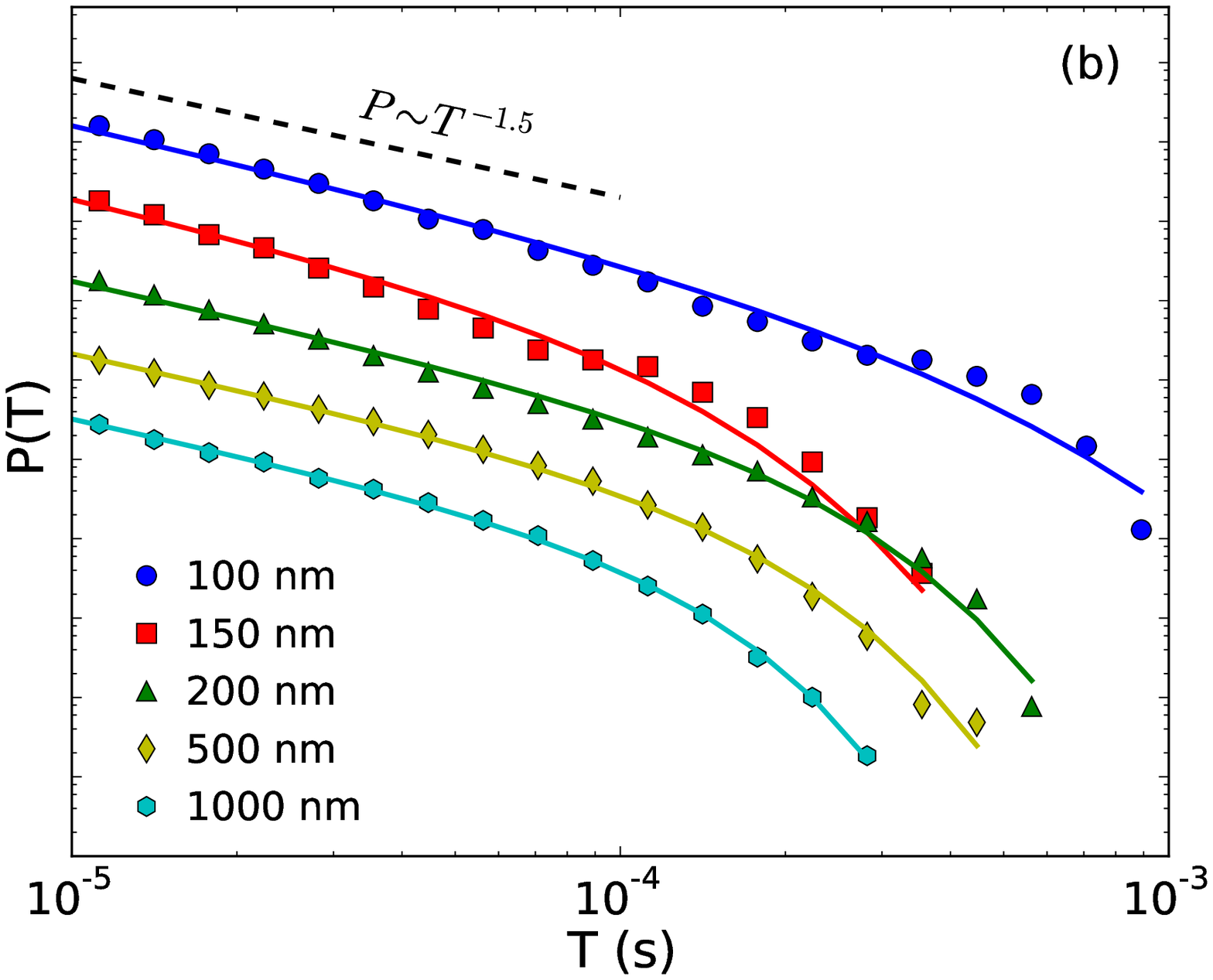} \\
\includegraphics[width=8.5cm]{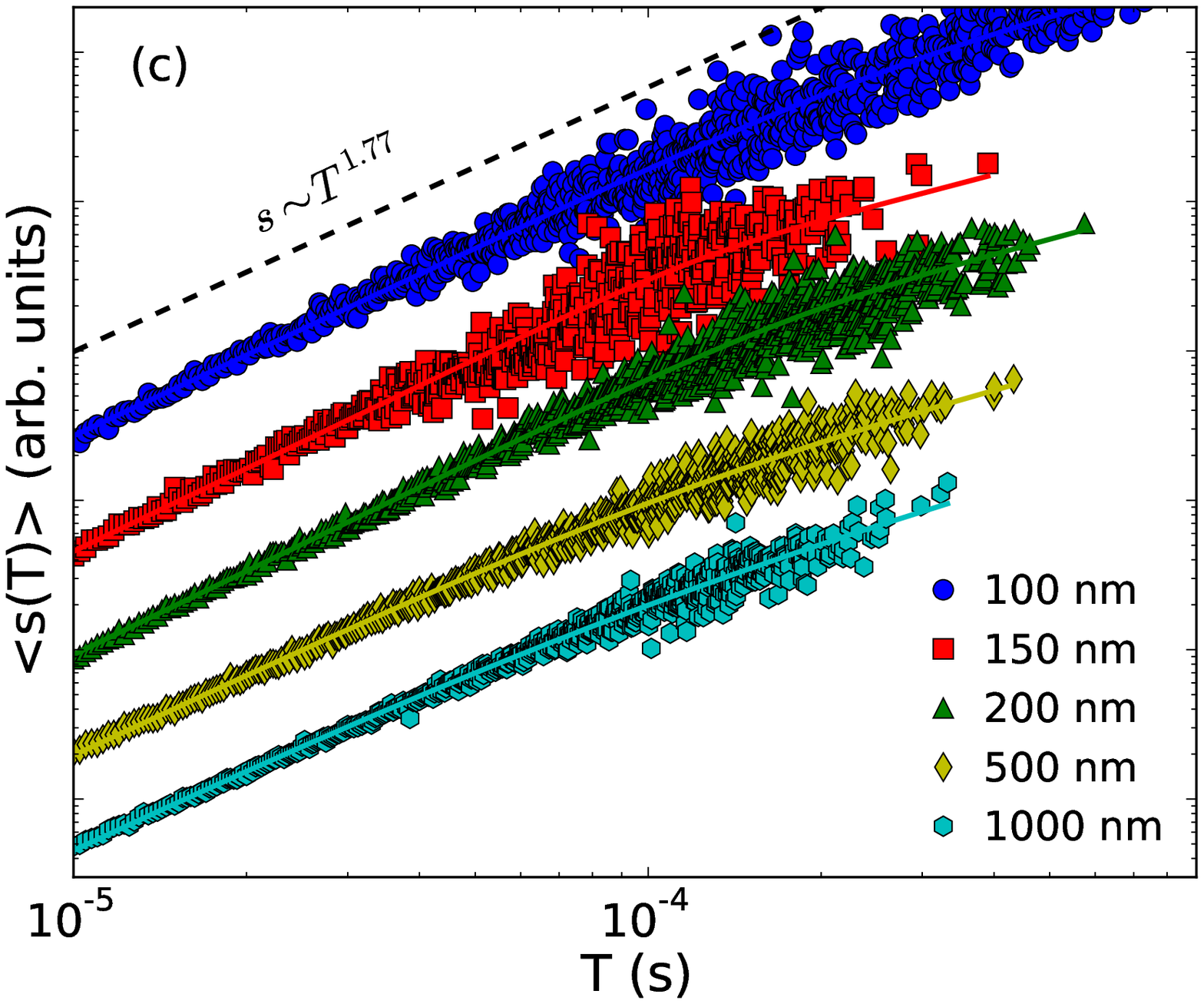}
\includegraphics[width=8.5cm]{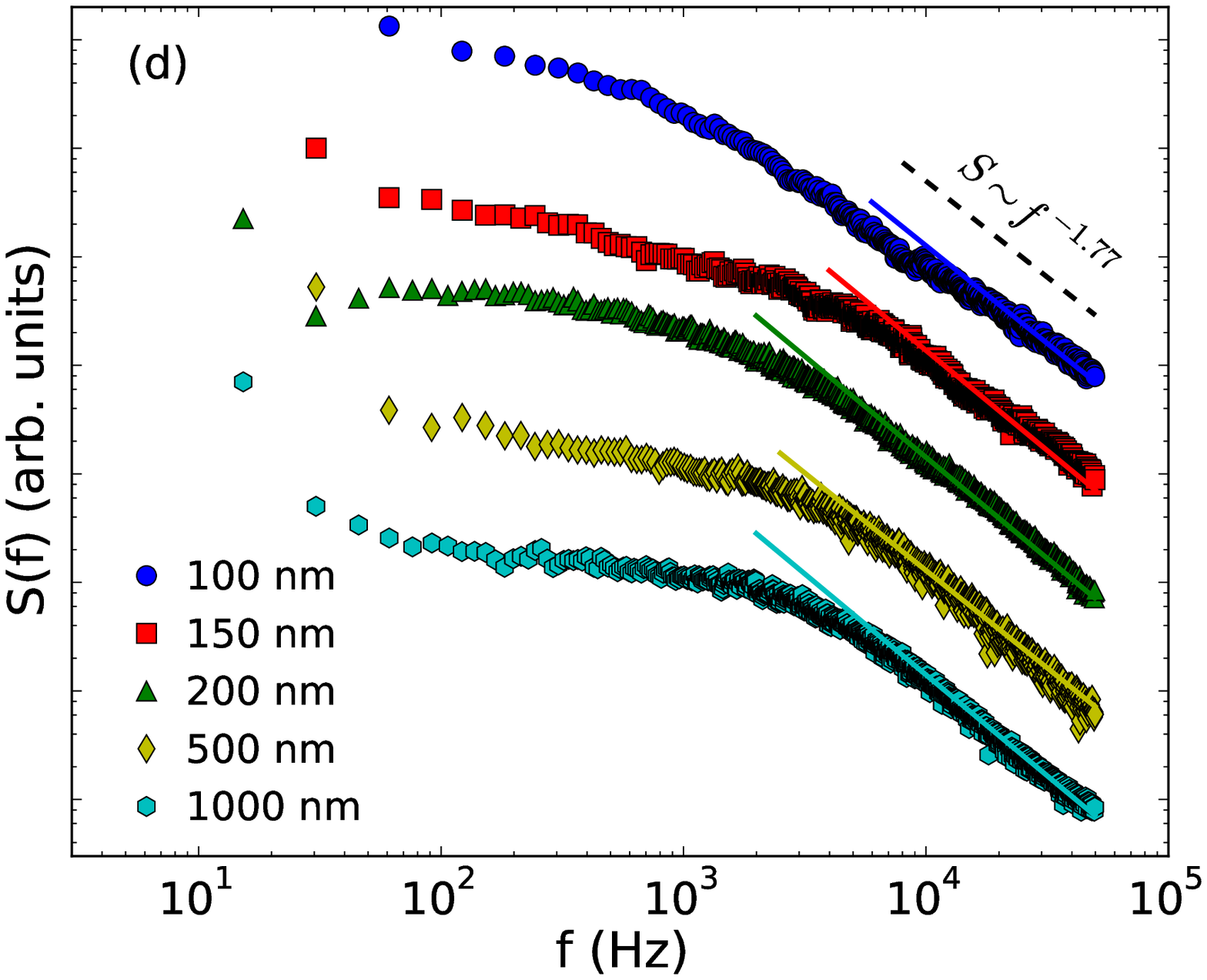} 
\vspace{-.3cm}\caption{(Color online) Traditional Barkhausen noise statistical properties measured for the FeSiB films with different thicknesses. (a) Distributions of Barkhausen avalanche sizes. The solid lines are cutoff-limited power-law fittings obtained using Eq.~(\ref{eq_01}), and for all FeSiB films, the fittings have exponent $\tau \sim 1.27$. (b) Similar plot for the distributions of Barkhausen avalanche durations, in which the solid lines are cutoff-limited power-law fittings obtained using Eq.~(\ref{eq_02}), and the fittings have exponent $\alpha \sim 1.5$. (c) Average avalanche size as a function of its duration, where the solid lines are cutoff-limited power-law fitting obtained using Eq.~(\ref{eq_03}), and the fitting have exponent $1/\sigma \nu z \sim 1.77$. (d) Finally, the power spectrum measured for the very same FeSiB films. To guide the eyes, the solid line are power-laws with slopes $\vartheta = 1/\sigma \nu z$ obtained respectively for each film. The best fit $\tau$, $\alpha$, and $1/\sigma \nu z$ exponents for FeSiB films with different thicknesses are given in Table \ref{exponents}. The distributions are shifted on vertical scale, to avoid superposition and make clearer the visualization.}
    \label{Fig_03} 
\end{figure*}

From the Barkhausen noise time series measured for our films, we perform the traditional statistical treatment employed for bulk materials. Figure~\ref{Fig_03} shows the distributions of Barkhausen avalanche sizes and durations, average avalanche as a function of its duration, and power spectrum obtained for the FeSiB films. In particular, the statistical properties are found to exhibit a cutoff-limited power-law behavior for all films, and they can be characterized by the exponents $\tau$, $\alpha$, $1/\sigma \nu z$, and $\vartheta$, respectively. 

We estimate the scaling exponents by fitting the experimental BN statistical properties considering the Eqs.~(\ref{eq_01})-(\ref{eq_03}). We determine the values of $\tau$, $\alpha$ and $1/\sigma \nu z$ jointly fitting the distributions $P(s)$, $P(T)$, and $\langle s(T) \rangle \, vs.\, T$ at the same time, respecting the scaling relation between the exponents $\tau$, $\alpha$, and $1/\sigma \nu z$, given in Eq.~(\ref{eq_04}). The fitted values of $n_s$, $n_T$, and $n_{ave}$ fall in the interval $1.2-3.0$. The results of the fits for the exponents are reported in Table~\ref{exponents} and also shown in Fig.~\ref{Fig_03}. The power spectrum has not been considered for the fitting procedure, however, we confirm the theoretical prediction of $1/\sigma \nu z = \vartheta$, indicating that the same scaling exponent can be employed for the relation between the average avalanche size and its duration as well as for the power spectrum at high frequencies~\cite{PRB62p11699, JMMM242p1085}, as can be verified in the same figure.
\begin{table}[!h]
\begin{center}
\caption{Values of $\tau$, $\alpha$ and $1/\sigma \nu z$ exponents for the experimental distributions measured for FeSiB amorphous ferromagnetic films with thicknesses of $100$, $150$, $200$, $500$, and $1000$ nm. The fits of $P(s)$, $P(T)$, and $\langle s(T) \rangle \, vs.\, T$ were performed simultaneously using Eqs.~(\ref{eq_01}), (\ref{eq_02}) and (\ref{eq_03}), respecting the scaling relation between the exponents, Eq.~(\ref{eq_04}).}
\label{exponents}
\begin{tabular}{cp{.25cm}cp{.25cm}cp{.25cm}c}
\hline \hline
Thickness (nm) && $\tau$ && $\alpha$ && $1/\sigma \nu z$ \\ \hline
$100$ && $1.30 \pm 0.04$ && $1.54 \pm 0.07$ && $1.80 \pm 0.07$\\
$150$ && $1.30 \pm 0.04$ && $1.55 \pm 0.07$ && $1.84 \pm 0.07$\\
$200$ && $1.28 \pm 0.03$ && $1.52 \pm 0.05$ && $1.86 \pm 0.05$\\
$500$ && $1.26 \pm 0.03$ && $1.47 \pm 0.04$ && $1.80 \pm 0.03$\\
$1000$ && $1.27 \pm 0.02$ && $1.50 \pm 0.03$ && $1.86 \pm 0.02$\\
\hline \hline
\end{tabular}
\end{center}
\end{table}

Based on these experimental statistical functions, the results show the scaling behavior of Barkhausen avalanches for the FeSiB films has similar scaling exponents, suggesting that they belong to a single universality class. These films do not show any noticeable dependence of the exponents on the field rate, in agreement with earlier findings for several amorphous samples~\cite{PRL84p4705, Resumao_BN}. Moreover, similarly to the features previously verified for polycrystalline films~\cite{PRE88p032811}, the exponents are independent on the film thickness, at least at this whole range of thickness, and present clear stability, despite the expected increase of the whole sample complexity with thickness, and large variations of the magnetic properties occuring between $200$ and $500$ nm. Thus, they corroborate the fact that the exponents are universal and independent of the microscopic details of each sample.

Regarding the actual values of the scaling exponents, the amorphous FeSiB ferromagnetic films with thicknesses between $100$ and $1000$ nm are characterized by scaling exponents $\tau\sim 1.27$, $\alpha \sim 1.5$, $1/\sigma \nu z \sim \vartheta \sim 1.77$.

The values of the scaling exponents are similar to that obtained for several bulk amorphous magnetic materials, $\tau=1.27\pm0.03$, $\alpha=1.5\pm0.1$ and $1/\sigma \nu z\sim \vartheta \sim 1.77$~\cite{PRL84p4705}, as well as to the ones previously reported by our group measured for amorphous FeCuNbSiB films in a wide range of thickness, $\tau \sim 1.25$ and $\alpha \sim 1.6$~\cite{PB384p144}, indicating a possible three dimensional magnetic behavior even at the smallest thicknesses. 

Besides, the measured exponent $\tau$ obtained here is distinct of the ones found in experimental works obtained with magneto-optical techniques for films with DWs dynamics governed by short-range interactions~\cite{NP3p547, JAP103p07D907}, since the studied films are thinner than $50$ nm and the known two-dimensional magnetic behavior is expected, and for amorphous films with variable thickness~\cite{JAP101p063903}. For the last case, the difference between the reported results and the ones presented here may be due to the limited penetration depth of the visible light in metals, around $10$ nm, restricting the probed depth of the material when considered magneto-optical techniques and defining the study to the magnetic properties of the film surface.

Several theoretical models have been proposed to explain the DWs dynamics and the BN statistical properties~\cite{Resumao_BN}. These works indicate the universality of the exponents and its dependence on the system dimensionality, although the predicted exponents vary according the theory~\cite{Resumao_BN}. Summarizing the theoretical predictions, for three-dimensional systems with the dynamics governed by long-range interactions, the scaling exponents are $\tau=1.50$, $\alpha=2.0$ and $1/\sigma \nu z=2$~\cite{PRL79p4669, PRB58p6353}, while for systems governed by short-range interactions and same dimensionality, $\tau=1.27$, $\alpha=1.5$ and $1/\sigma \nu z=1.77$~\cite{PRL79p4669, PRB58p6353, PRE69p026126, PRE64p066127}. On the other side, for two-dimensional systems, although there is not a complete agreement between theoreticians on the real values, the models indicate $\tau\sim1.33$, $\alpha\sim1.5$ and $1/\sigma \nu z\sim 1.5$ for the long-range interaction problem~\cite{JSMP08020, IEEETM46p228, PRL84p1316, PRB89p104402}, while $\tau\sim1.06$ for the short-range interaction one~\cite{PRE69p026126, PRE64p066127, PRB89p104402}. In the last case, $\alpha$ and $1/\sigma \nu z$ are still not predicted.

In particular, we identify that the scaling exponents measured here for the FeSiB films are in quantitative agreement with the exponent values predicted by the model proposed by P.~Cizeau, S.~Zapperi, G.~Durin, and H.~E.~Stanley (CZDS model), if dipolar interactions are neglected~\cite{PRL79p4669, PRB58p6353}, $\tau = 1.27$, $\alpha = 1.5$, and $1/\sigma \nu z=1.77$, and by the model originally introduced by J.~S.~Urbach, R.~C.~Madison, and J.~T.~Marker (UMM model)~\cite{PRL75p276} and investigated by S.~L.~A.~de Queiroz~\cite{PRE69p026126, PRE64p066127}, $\tau = 1.27$. More than an experimental evidence to show that the CZDS and UMM models can be extended to describe the BN statistical properties in films, the scaling exponents also corroborate the universality class of amorphous alloys under stress, related to short-range interactions, as proposed in Ref.~\cite{PRL84p4705}. 

\begin{figure}[!] 
    \includegraphics[width=8.5cm]{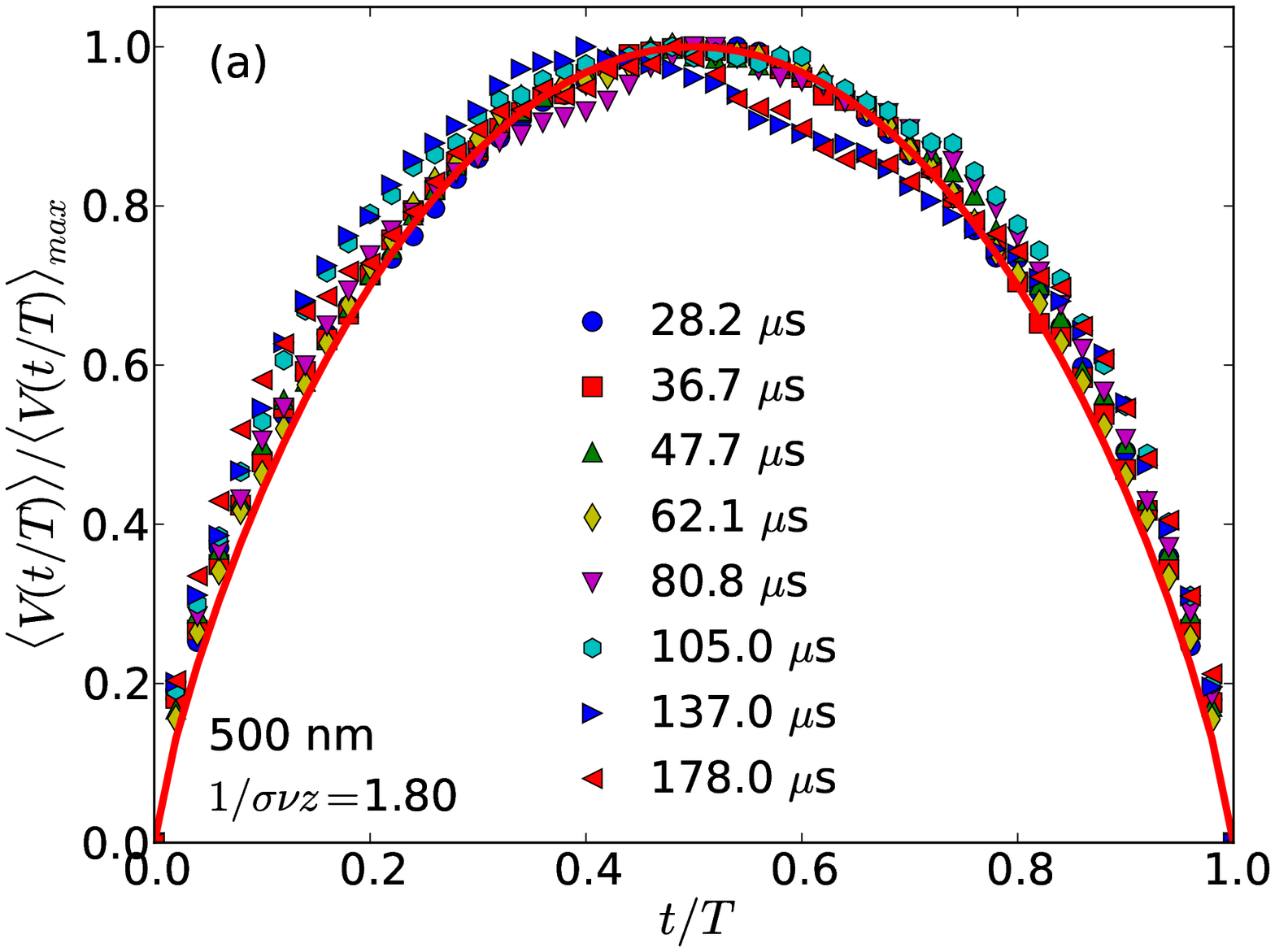} \\
    \includegraphics[width=8.5cm]{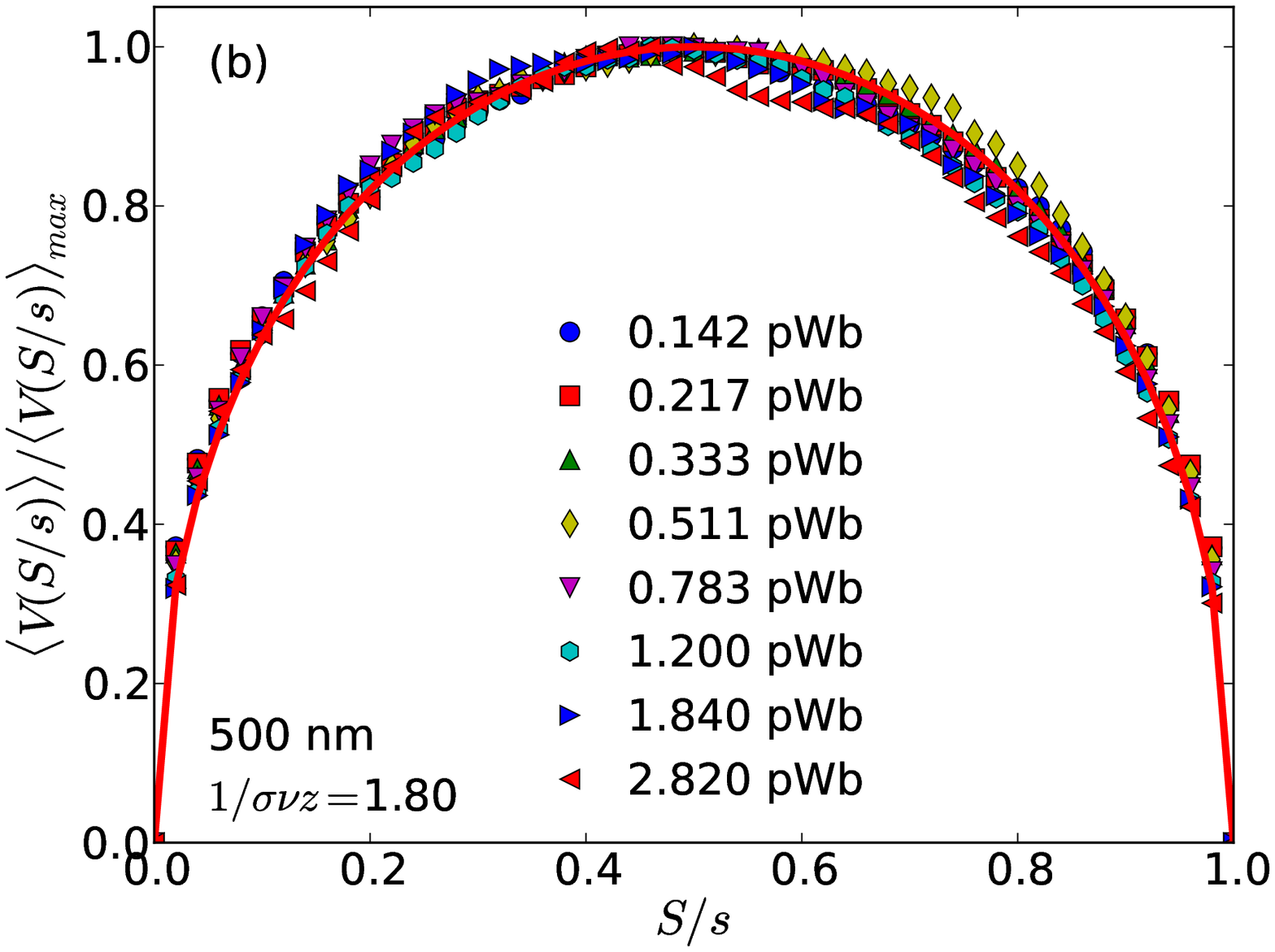} 
\vspace{-.3cm}\caption{(Color online) Experimental average avalanche shapes measured for the $500$ nm-thick FeSiB film and theoretical predictions for the average avalanche shapes for non-mean-field systems. (a) Average temporal avalanche shape for different avalanche durations $T$, rescaled to unit height and duration. Different symbols correspond to different durations of the avalanches, while the solid line is the theoretical predicion according to Eq.~(\ref{eq_06}) with the best fit $1/\sigma \nu z$ exponent measured from the  curve of the average avalanche size as a function of its duration. (b) Similar plot for the average avalanche shape for different avalanche sizes $s$, rescaled to unit height and size. Symbols correspond to distinct sizes of the avalanches, and the solid line is the theoretical prediction according to Eq.~(\ref{eq_07}) with the very same best fit $1/\sigma \nu z$ exponent.} 
    \label{Fig_04}
\end{figure}

Finally, as a refined tool to characterize materials and test universality classes, we focus on the measurement of the average avalanche shape. Figure~\ref{Fig_04} shows the average shapes measured for different avalanche durations and sizes for the $500$ nm-thick FeSiB film, as a representative example of the experimental results obtained for the studied films. The theoretical predictions for the scaling form for the average temporal avalanche shapes and shape of avalanches of a given size, given by Eqs.~(\ref{eq_06}) and (\ref{eq_07}), respectively, obtained when the best fit exponent $1/\sigma \nu z$ is considered are also shown in Fig.~\ref{Fig_04}. In the scaling regime verified in the curve of the average avalanche size as a function of its duration, this film is characterized by the exponent $1/\sigma \nu z = 1.80 \pm 0.03$. Notice the striking quantitative agreement between experiment and theoretical predicions, including three important features: symmetric shapes, the exponent $1/\sigma \nu z$, and the exact form of the average avalanche shapes.

Regarding the symmetry of the shapes, the average shapes are not characterized by any evident leftward asymmetry, as observed in amorphous and polycrystalline bulk samples~\cite{Resumao_BN}. Doubts about asymmetry of the shapes were resolved when eddy currents were show to be responsible for the it~\cite{NP1p46}, i. e., the asymmetry is devoted to the non instantaneous response of the eddy field to the wall displacement, a direct signature of the negative effective mass associated with the domain wall moving under the action of the external field~\cite{Resumao_francesca}. Besides, it has been verified for bulk samples that the asymmetry depends on the avalanche duration and encodes important information on the characteristic time of the underlying dynamics. Experimentally, the asymmetry is dependent on the material parameters, as the magnetic permeability $\mu$ conductivity $\sigma$, as well as geometrical dimensions of the sample, as the thickness. Here, by employing films with intermediate thickness, the characteristic timescale for relaxation~\cite{NP1p46} is of $\sim$ns, value smaller than the range of the avalanche durations, above $\sim \mu$s. For this reason, the domain wall dynamics seems to be not affected by eddy current effects, resulting in symmetric average shapes, undistorted by eddy currents.

On the form, average avalanche shape depends on the universality class of the avalanche dynamics~\cite{NC4p2927}. In mean-field systems, such as polycrystalline films with intermediate thicknesses~\cite{NP7p316, PRE88p032811}, with $1/\sigma \nu z \sim 2.0$, the average temporal avalanche shape is known to be given by an inverted parabola, while the average shape for different sizes is given by a semicircle. Here, the average avalanche shapes are in quantitative agreement with theoretical predictions proposed in Ref.~\cite{NC4p2927}, indicating that the best-fit exponent obtained through the average avalanche size as a function of its duration, and the parabola and semicircle with corrections in which the $1/\sigma \nu z$ is considered are appropriate to describe these amorphous films. Thus, we provide experimental support for the validity of a general scaling form for the average avalanche shapes for non-mean-field systems.

After all, considering power-laws and shapes, we interpret the concordance between experiment and theory as a clear indication that the FeSiB amorphous films, within this range of thickness, present a typical three-dimensional magnetic behavior with predominant short-range elastic interactions governing the DWs dynamics. 
%\begin{figure}[!h] 
%    \includegraphics[width=8.5cm]{Fig_as_500nm_08.eps} \\
%\vspace{-.3cm}\caption{(Color online) Average avalanche size as a function of its duration for the FeSiB films with different thicknesses. The plots are shifted on vertical scale, to avoid superposition and make clearer the visualization. The solid lines are cutoff-limited power-law fitting obtained using Eq.~(\ref{eq_03}). The best fit $1/\sigma \nu z$ exponents are given in Table \ref{exponents}. For all Permalloy films, the fitting have exponent $1/\sigma \nu z \sim 1.77$.} 
%    \label{Fig_08} 
%\end{figure}
%\begin{figure}[!h] 
%    \includegraphics[width=8.5cm]{Fig_as_500nm_10.eps} 
%\vspace{-.3cm}\caption{(Color online) Average avalanche size as a function of its duration for the FeSiB films with different thicknesses. The plots are shifted on vertical scale, to avoid superposition and make clearer the visualization. The solid lines are cutoff-limited power-law fitting obtained using Eq.~(\ref{eq_03}). The best fit $1/\sigma \nu z$ exponents are given in Table \ref{exponents}. For all Permalloy films, the fitting have exponent $1/\sigma \nu z \sim 1.77$.} 
%    \label{Fig_10} 
%\end{figure}

\section{Conclusion}  
\label{Conclusion}

In summary, in this paper we investigate the statistical properties of the Barkhausen noise in amorphous ferromagnetic films in a wide range of thicknesses, from $100$ to $1000$ nm. From Barkhausen noise time series measured with the traditional inductive technique, we perform a wide statistical analysis and establish the scaling exponents $\tau$, $\alpha$, $1/\sigma \nu z$, and $\vartheta$, as well as we also focus on the average shape of the avalanches. 

Through this wide statistical analysis, we group the amorphous films with distinct thicknesses in a single universality class, characterized by scaling exponents $\tau\sim 1.27$, $\alpha \sim 1.5$, $1/\sigma \nu z \sim \vartheta \sim 1.77$. The measured scaling exponents are similar to that obtained for several bulk amorphous magnetic materials and amorphous films in a wide range of thickness, as well as are in quantitative agreement with the predictions of two theoretical models: CZDS model, if dipolar interactions are neglected~\cite{PRL79p4669, PRB58p6353}, and the UMM model~\cite{PRL75p276, PRE69p026126}. Our films are thinner than ribbons and sheets~\cite{PRL84p4705}, known to exhibit three-dimensional magnetic behavior, but thicker than previously studied two-dimensional films~\cite{PRL84p5415, IEEETM36p3090, PRL90p0872031, JMMM310p2599, JAP93p6563, NP3p547, JAP103p07D907, SSC150p1169, JAP109p07E101, PRB83p060410R}. We interpret these results as a clear evidence that these amorphous films, within this thickness range, present a typical three-dimensional magnetic behavior with predominant short-range elastic interactions governing the domain wall dynamics.

In addition, when we consider the average avalanche shape, experimental results are in quantitative agreement with theoretical predictions ~\cite{NC4p2927}. In our amorphous films, we find striking symmetric shapes, undistorted by eddy currents, in which the average avalanche shape for different durations and sizes are respectively described by a parabola and a semicircle with corrections in which the exponent $1/\sigma \nu z$ is considered. Besides we verify that the average shape correpond to a powerful tool do characterize classes, and provide experimental support for the validity of a general scaling form for the average avalanche shapes for non-mean-field systems.

These results obtained for the Barkhausen noise statistical properties for amorphous films in a wide range of thicknesses, associated to the ones measured for polycrystalline films previously reported in Ref.~\cite{PRE88p032811}, provide a further insight on the DWs dynamics in ferromagnetic films and the role of structural character on the scaling behavior in the BN statistical properties. In particular, we verify that materials can be classified in different universality classes, and confirm that the classes proposed in Ref.~\cite{PRL84p4705} for bulk samples can be extended for films, even with thickness down to $100$ nm, corroborating the link between material microstructure and the BN statistical properties is still valid in systems with reduced dimensions. The next step here resides basicaly in extending the studies to wider ranges of film thicknesses. Experiments and analyses on the domain wall dynamics in thinner films are currently in progress.

\begin{acknowledgments} 
F.B.\ would like to thank L.\ Laurson for the fruitful discussions. The research is supported by the Brazilian agencies CNPq (Grants No.~$471302$/$2013$-$9$, No.~$310761$/$2011$-$5$, No.~$476429$/$2010$-$2$, No.~$555620$/$2010$-$7$, No.~$307951$/$2009$-$0$, and No.~$482735$/$2009$-$0$), CAPES, FAPERJ (No.~E-$26$/$102.943$/$2008$, and No.~E-$26$/$112.697$/$2012$), and FAPERN (Grant Pronem No.~$03/2012$), Progetto Premiale MIUR-INRIM ``Nanotecnologie per la metrologia elettromagnetica'', and MIUR-PRIN 2010-11 Project2010ECA8P3 ``DyNanoMag''. M.A.C.\ and F.B.\ acknowledge financial support of the INCT of Space Studies.
\end{acknowledgments} 

%\newpage
\bibliographystyle{References_h-physrev3}
\bibliography{References_FBohn}
%\begin{thebibliography}{999} 
%\bibitem{PZ20p401}H.~Barkhausen, Phys.\ Z.\ {\bf 20}, 401 (1919). 
%\end{thebibliography}

\end{document}